\documentclass{fizik}
\usepackage{multicol,epsfig}

\vol{27} 
\pyear{2003} 
\received{\today}

\author[RABENSTEIN, AVERIN]{

\textbf{Kristian RABENSTEIN and Dmitri V. AVERIN} \\ 
\textit{Department of Physics and Astronomy, Stony Brook 
University, SUNY}\\
\textit{Stony Brook, NY 11794, U.S.A.}\\

}

\title{Decoherence in two coupled qubits}

\begin{document}
\maketitle

\begin{abstract}

We have developed quantitative description of quantum coherent 
oscillations in the system of two coupled qubits in the presence 
of weak decoherence that in general can be correlated between the 
two qubits. It is shown that in the experimentally realized 
scheme of excitation of the oscillations, their waveform is 
not very sensitive to the magnitude of decoherence correlations. 
Modification of this scheme into potentially useful probe of the 
degree of decoherence correlations at the two qubits is suggested. 

\end{abstract}

\section{Introduction}

Despite the large number of successful demonstrations of quantum 
coherent oscillations in individual \cite{b1,b2,b3,b4,b5,b6} and 
coupled \cite{b7,b8} Josephson-junction qubits, quantitative 
understanding of the details of these oscillations is so far quite 
limited. One of the most important open problems is decoherence,  
many aspects of which still remain to be understood. The purpose 
of this work is to develop theoretical description of decoherence 
in dynamics of a system of two coupled qubits. The approach we are 
using is based on the evolution equation for the density matrix in 
the Markovian approximation that is standard for description of 
weak decoherence. Although there are indications from the 
single-qubit experiments that more sophisticated approaches are 
needed for quantitatively accurate description of decoherence, 
the result we obtain within the simple scheme can be useful as the 
benchmark for more elaborate models.   

Motivation for studying decoherence in coupled qubits is provided 
by the recent experiment \cite{b7} with two 
coupled charge qubits, where it was found that the decoherence 
rate for quantum coherent oscillations in two qubits at the 
optimal bias point is with good accuracy factor-of-4 larger than 
the decoherence rate in effectively decoupled qubits. An 
interesting question for theory is whether this factor-of-4 
increase of the decoherence rate is a numerical coincidence, or it 
reflects some basic property of the decoherence mechanisms in 
charge qubits. As will become clear from the discussion below, the 
theory developed in this work favors ``numerical coincidence'' 
point of view. Other aspects of decoherence in coupled qubits has 
been studied before numerically in \cite{b14,b15,b16}. 

In general, it is well understood that decoherence rates of 
different states of two coupled qubits can be quite different if 
the random forces created by the qubit environments responsible for 
decoherence are completely or partially correlated at the two qubits. 
Most importantly, in the case of complete correlation, the qubit 
system should have a ``decoherence-free subspace'' (DFS) spanned by the 
states $|01\rangle$, $|10\rangle$ \cite{b9,b10,b11}, since completely 
correlated external environments can not distinguish these states. 
In contrast, the subspace spanned by $|00\rangle$ and $|11\rangle$ 
experiences decoherence that is made stronger by the correlations 
between environmental forces acting on the two qubits. So the role 
of the quantitative theory in description of decoherence in the 
dynamics of coupled qubits is to see to what extent subspaces 
with different decoherence rates participate in the qubit 
oscillations for different methods of their excitation.

\section{The model and environmental correlations}  

The Hamiltonian of the system of two qubits coupled directly by 
interaction between the basis-forming degrees of freedom (i.e., 
electrostatic interaction through finite coupling capacitance 
for charge qubits, or magnetic interaction for flux qubits) is: 
\begin{equation} 
H_0 = \sum_{j=1,2} (\varepsilon_j \sigma_{z}^{(j)} + 
\Delta_j \sigma_{+}^{(j)}+\Delta_j^* \sigma_{-}^{(j)}) +\nu 
\sigma_{z}^{(1)} \sigma_{z}^{(2)} \, ,
\label{e1} \end{equation}  
where $\sigma$'s denote the Pauli matrices, $\nu$ is the qubit 
interaction energy, $\Delta_j$ is the tunnel amplitude and 
$\varepsilon_j$ is the bias of the $j$th qubit. Four energy levels 
of the Hamiltonian (\ref{e1}) are shown schematically in Fig.\ 1 
as functions of the common bias $\bar{\varepsilon} \equiv 
\varepsilon_1=\varepsilon_2$ of the two qubits. In this work, we 
will consider quantum coherent oscillations in the qubits biased 
at the ``co-resonance'' point \cite{b7}, where $\varepsilon_1= 
\varepsilon_2 =0$. Such bias conditions are optimal for the 
oscillations. 

It can be shown explicitly that the occupation probabilities 
of the qubit basis states (that are of interest for us) 
are insensitive to the phases of the qubit tunnel amplitudes 
$\Delta_j$, so without the loss of generality we will assume 
that $\Delta_j$'s are real. The Hamiltonian (\ref{e1}) at the 
co-resonance reduces then to 
\begin{equation} 
H_0 = \sum_{j=1,2} \Delta_j \sigma_{x}^{(j)} +\nu \sigma_{z}^{(1)} 
\sigma_{z}^{(2)} \, .
\label{e2} \end{equation}  
In the basis composed of eigenstates of the $\sigma_{x}^{(j)}$ 
operators, the Hamiltonian (\ref{e2}) can be diagonalized easily. 
Eigenenergies and eigenstates are: 
\begin{eqnarray} 
& E_1=\Omega, & |\psi_1\rangle =\frac{1}{2} [(\gamma+\eta) 
(|00\rangle +|11\rangle) +(\gamma-\eta) (|01\rangle +|10\rangle) 
] , \nonumber \\
& E_2=-\Omega, & |\psi_2\rangle =\frac{1}{2} [(\eta-\gamma) 
(|00\rangle +|11\rangle) +(\gamma+\eta) (|01\rangle +|10\rangle) 
] , \label{e3}  \\
& E_3=\epsilon, & |\psi_3\rangle =\frac{1}{2} [(\alpha+\beta) 
(|00\rangle -|11\rangle) +(\alpha-\beta) (|10\rangle -|01\rangle) 
] , \nonumber \\
& E_4=-\epsilon, & |\psi_4\rangle =\frac{1}{2} [(\beta-\alpha) 
(|00\rangle -|11\rangle) +(\alpha+\beta) (|10\rangle -|01\rangle) 
] , \nonumber 
\end{eqnarray}
where 
\[ \Omega=(\Delta^2+\nu^2)^{1/2}\, , \;\;\; \epsilon=(\delta^2+ 
\nu^2)^{1/2}\, , \;\;\; 
\alpha, \beta = \frac{1}{\sqrt{2}}(1\pm \frac{\delta} 
{\epsilon})^{1/2} \, , \;\;\; \eta,\gamma= \frac{1}{\sqrt{2}} 
(1\pm \frac{\Delta}{\Omega})^{1/2} \, ,\]
and $\Delta\equiv \Delta_1+\Delta_2$, $\delta \equiv \Delta_1- 
\Delta_2$. The states $|kl\rangle$ with $\{ k,l\} =\{ 0,1\}$ in 
Eqs.~(\ref{e3}) are the eigenstates of the operators $\sigma_{z 
}^{(1,2)}$ in the natural notations: $\sigma_{z}^{(1)} 
|kl\rangle = (-1)^k|kl\rangle$ and $\sigma_{z}^{(2)}|kl\rangle = 
(-1)^l|kl\rangle$. 

\begin{figure}[htb]
\setlength{\unitlength}{1.0in}
\begin{picture}(5.0,3.1) 
\put(2.0,-.05){\epsfxsize=2.3in\epsfbox{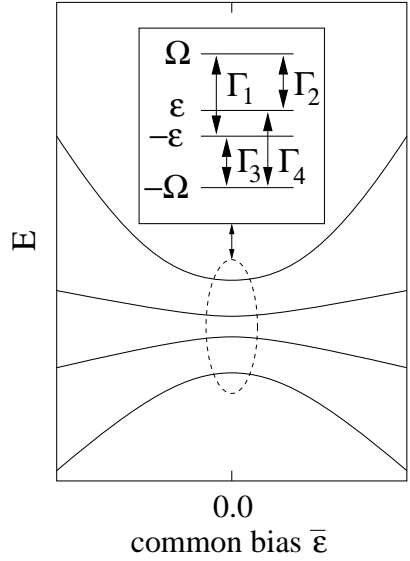}}
\end{picture}
\caption{ Schematic structure of the energy levels of the two 
coupled qubits as functions of the common bias of the qubits. 
The inset shows the diagram of the decoherence-induced 
transitions between the levels at ``co-resonance'' point 
where the bias vanishes.}
\end{figure}

We assume that external environments responsible for the 
decoherence couple to the basis-forming degrees of freedom of 
the qubits. The interaction Hamiltonian is then: 
\begin{equation} 
H_i= \sum_{j=1,2} \xi_j(t) \sigma_{z}^{(j)}  \, . 
\label{e4} \end{equation}  
The random forces $\xi_{1,2}(t)$ acting on the qubits are in 
general correlated. To describe the weakly dissipative dynamics 
of the system in the basis of states (\ref{e3}) induced by the 
interaction (\ref{e4}) with the reservoirs, we use the standard 
equation for the evolution of the qubit density matrix $\rho$ 
in the interaction representation (see, e.g., \cite{b12}): 
\begin{equation} 
\dot{\rho}=-\int_{-\infty}^t d\tau \langle [H_i(t),[H_i(\tau), 
\rho]] \rangle \, ,  
\label{e5} \end{equation}  
where angled brackets denote averaging over the reservoirs. 
Proceeding in the standard way, we keep in Eqn.\ (\ref{e5}) only 
terms that do not oscillate as functions of time with large 
frequencies on the order of eigenenergies of the system, and 
therefore lead to changes in $\rho$ that accumulate over time. 
Equations for the matrix elements $\rho_{nm}$, $n,m=1,...,4$, of 
$\rho$ in the basis (\ref{e3}) are transformed then as follows: 
\begin{eqnarray} 
\dot{\rho}_{nm} & = & \sum_{j,j'=1,2}\big[-\rho_{nm} 
(\sigma_{mm}^{(j)}-\sigma_{nn}^{(j)})(\tilde{F}_{jj'}^*(0)
\sigma_{mm}^{(j')} -\tilde{F}_{jj'}(0)\sigma_{nn}^{(j')})  
\nonumber \\ 
& & -\rho_{nm}\sum_{k}(\sigma_{nk}^{(j)}\sigma_{kn}^{(j')} 
\tilde{F}_{jj'}(\epsilon_{n}-\epsilon_{k})
+\sigma_{mk}^{(j')}\sigma_{km}^{(j)}\tilde{F}_{jj'}^*
(\epsilon_{m} -\epsilon_{k}))
\nonumber \\
& & +\delta_{nm}\sum_{k}\rho_{kk}(\sigma_{nk}^{(j')}
\sigma_{kn}^{(j)}\tilde{F}_{jj'}(\epsilon_{k}-\epsilon_{n})+
\sigma_{nk}^{(j)} \sigma_{kn}^{(j')}\tilde{F}_{jj'}^*
(\epsilon_{k}-\epsilon_{n})
\nonumber \\
& & + \sum_{(k,l)}\rho_{kl}\sigma_{nk}^{(j')}
\sigma_{lm}^{(j)} (\tilde{F}_{jj'}(\epsilon_{l}-\epsilon_{m})+
\tilde{F}_{j'j}^*(\epsilon_{l}-\epsilon_{m})) \big].  
\label{e6} \end{eqnarray}
Here $\sigma_{nm}^{(j)}$ denote the matrix elements $\langle 
n|\sigma_z^{(j)}|m\rangle$, the last sum is taken over the pairs 
$(k,l)$ of states that satisfy the ``resonance'' condition: 
\[ \epsilon_{k}-\epsilon_{l}= \epsilon_{n}-\epsilon_{m} \, , \;\;\; 
(k,l) \neq (n,m) \, , \]
and 
\[ \tilde{F}_{jj'}(\omega)=\int_{0}^{\infty}dt\langle \xi_{j}(t)
\xi_{j'}(0)\rangle e^{i\omega t} \, . \]

The first term in Eqn.~(\ref{e6}) represents ``pure dephasing'' that 
exists when the system operators that couple it to the environment 
have non-vanishing average values in the eigenstates. As one can 
see explicitly from Eqs.\ (\ref{e3}), the average values 
$\sigma_{nn}^{(j)}$ of $\sigma_z^{(j)}$ are vanishing in all states, 
so that there is no pure dephasing term in the evolution of the 
density matrix in our case. The fact that $\sigma_{nn}^{(j)}$ are 
vanishing can be related to the vanishing slope of the system energies 
with respect to variations of the bias in the vicinity of the 
co-resonance point -- see Fig.~1. Since all coefficients in the 
eigenfunctions (\ref{e3}) are real, the matrix elements 
$\sigma_{nm}^{(j)}$ are also real. For real $\sigma_{nm}^{(j)}$, 
imaginary parts of the noise correlators $\tilde{F}_{j'j}$ in the 
second term on the right-hand-side of Eqn.~(\ref{e6}) represent the 
decoherence-induced shifts of the system energy levels. These 
shifts do not affect decoherence and we will neglect them in 
our discussion. With these simplifications, Eqn.~(\ref{e6}) takes 
the form
\begin{eqnarray} 
\dot{\rho}_{nm} & = & \sum_{j,j'=1,2}\big[ -(\rho_{nm} /2) 
\sum_{k}(\sigma_{nk}^{(j)}\sigma_{kn}^{(j')} \mbox{Re} F_{jj'} 
(\epsilon_{n}-\epsilon_{k})+\sigma_{mk}^{(j')}\sigma_{km}^{(j)} 
\mbox{Re} F_{jj'} (\epsilon_{m} -\epsilon_{k}))
\nonumber \\
& & +\delta_{nm}\sum_{k}\rho_{kk} \sigma_{nk}^{(j')}
\sigma_{kn}^{(j)}\mbox{Re} F_{jj'}(\epsilon_{k}-\epsilon_{n})  
+ \sum_{(k,l)}\rho_{kl}\sigma_{nk}^{(j')} \sigma_{lm}^{(j)} 
F_{jj'}(\epsilon_{l}-\epsilon_{m}) \big],  
\label{e7} \end{eqnarray}
where 
\begin{equation} 
F_{jj'}(\omega)=\int_{-\infty}^{\infty}dt\langle \xi_{j}(t)
\xi_{j'}(0)\rangle e^{i\omega t} \, . 
\label{e8} \end{equation} 

The function $F_{12}$ characterizes correlations between the 
environmental forces  acting on the two qubits. For instance, 
if the two qubits interact with different environments and 
$\xi_1$, $\xi_2$ are uncorrelated, $F_{12}=0$, whereas 
$F_{12}=F_{11}=F_{22}$, if the qubits are acted upon by the 
force produced by one environment coupled equally to the 
two qubits. While the correlators 
$F_{11}$ and $F_{22}$ are necessarily real, $F_{12}$ can be 
imaginary, and $F_{21}^*=F_{12}$. Non-vanishing imaginary part of 
$F_{12}$ corresponds to the non-vanishing commutator $[\xi_1, 
\xi_2]$ and implies that the two qubits are coupled to the two 
non-commuting variables of the same reservoir. While this is 
probably not very likely for qubits with the basis-forming 
degrees of freedom of the same nature (which in a typical situation 
should be coupled to the same set of environmental degrees of 
freedom), the non-vanishing $\mbox{Im}F_{12}$ should be typical if 
the qubits have different basis-forming variables. Using the 
spectral decomposition of the correlators $F_{jj'}(\omega)$ and 
Swartz inequality, one can prove (similarly to what is done in a 
different context of linear quantum measurements \cite{b13}) that 
for arbitrary stationary reservoirs the correlators satisfy the 
inequality that imposes the constraint on $F_{12}(\omega)$: 
\begin{equation} 
F_{11}(\omega)F_{22}(\omega)\geq |F_{12}(\omega)|^2 \, .
\label{e9} \end{equation} 
If the reservoirs are in equilibrium at temperature $T$, the 
correlators satisfy also the standard detailed balance relations: 
\begin{equation} 
F_{jj'}(-\omega) =e^{-\frac{\omega}{T}} F_{j'j}(\omega) \, . 
\label{e10} \end{equation} 

Equation (\ref{e7}) with the noise correlators (\ref{e8}) 
govern weakly dissipative time evolution of the two coupled 
qubits in a generic situation. Below we use them to determine
decoherence properties of quantum coherent oscillations of the 
qubits. Before doing this, however, we would like to briefly 
discuss applicability of our approach to realistic 
Josephson-junction qubits. As we saw above, one of the main 
features of Eqn.~(\ref{e7}) is that the pure dephasing terms 
disappear at the co-resonance point and remaining decoherence 
is related to transitions between the energy eigenstates. This 
implies that within the approach based on Eqn.~(\ref{e7}), the 
decoherence rates are on the order of half of the transition 
rates, whereas experiments with charge qubits (see, e.g., 
\cite{b4}) indicate that decoherence rates are larger than the 
transition rates even at the optimum bias point when the  
pure-dephasing terms should disappear. Apparently, this is related 
to the low-frequency charge noise \cite{b4,b17} that is coupled to 
qubit strongly enough for the lowest-order perturbation theory in 
coupling (\ref{e5}) to be insufficient. This implies that the 
theory presented in this work might be only qualitatively correct 
for realistic charge qubits, for which one should develop more 
accurate non-perturbative description of the low-frequency noise 
to achieve quantitative agreement with experiments. Our simple 
perturbative approach, however, should be applicable to flux 
qubits, where the low-frequency noise should not be as strong as 
in the charge qubits.

\section{Quantum coherent oscillations in coupled qubits}

One of the most direct ways of excitation of quantum coherent 
oscillations  in individual or coupled qubits that will be 
discussed in this work is based on the abrupt variation of the 
bias conditions \cite{b1,b7}. If the qubits are initially 
localized in one of their basis states, e.g. $|00\rangle$, and 
abrupt variation of the bias brings them to the co-resonance point, 
the probabilities for the qubits to be in other basis states 
start oscillating with time. In the simplest detection scheme 
(realized, for instance, in experiment \cite{b7}) the probability 
for each qubit to be in the state $|1\rangle$ is measured 
independently of the state of the other qubit. Quantitatively, 
these probabilities $p_j$ are obtained from the projection 
operators $P_j$: 
\[ p_j=\mbox{Tr}\{\rho P_j \}\, ,\;\;\; P_1= \sum_{k=1,2} 
|1k\rangle \langle k1| \, , \;\;\; P_2= \sum_{k=1,2} 
|k1\rangle \langle 1k|  \, . \] 
Finding explicitly the matrix elements of $P_j$ from the 
wavefunctions (\ref{e3}), one gets: 
\begin{eqnarray} 
p_{1}(t)=\frac{1}{2}+ (\alpha\eta+\beta\gamma) \mbox{Re} 
[e^{-i\omega_- t}(\rho_{42}(t)-\rho_{13}(t))] +(\alpha \gamma
- \beta\eta)\mbox{Re} [e^{-i\omega_+ t} (\rho_{14}(t)+
\rho_{32}(t))] \, , \\
p_{2}(t)=\frac{1}{2}-(\alpha\gamma + \beta\eta) \mbox{Re} 
[e^{-i\omega_- t} ( \rho_{13}(t)+\rho_{42}(t))] 
+(\alpha\eta-\beta\gamma) \mbox{Re} [e^{-i\omega_+ t}
(\rho_{14}(t)-\rho_{32}(t))] \, ,
\label{e11} \end{eqnarray} 
where $\omega_{\pm} \equiv \Omega \pm \epsilon $, and as in 
the Eqn.~(\ref{e7}), the matrix elements of the density matrix 
are taken in the interaction representation. Equations (\ref{e11}) 
and (\ref{e7}) show that the waveform of the coherent oscillations 
in coupled qubits is determined by the time evolution of the two 
pairs of the matrix elements of $\rho$: 
\begin{equation}
\dot{\rho}_{13}=-\Gamma_{13}\rho_{13}+u_-\rho_{42} \, , 
\;\;\; \dot{\rho}_{42}=-\Gamma_{42}\rho_{42}+u_+\rho_{13} \, ,
\label{e12} \end{equation} 
\begin{equation}
\dot{\rho}_{14}=-\Gamma_{14}\rho_{14}+v_- \rho_{32} \, , 
\;\;\; \dot{\rho}_{32}=-\Gamma_{32}\rho_{32}+v_+ \rho_{14} \, . 
\label{e13} \end{equation} 

The decoherence rates in these equations are determined by the 
rates of transitions between different energy eigenstates: 
\begin{eqnarray}
\Gamma_{13}=\frac{1}{2}\big(\Gamma_1^{(+)}+ \Gamma_2^{(-)}+ 
\Gamma_2^{(+)}+\Gamma_4^{(+)}\big) \, , \;\;\; \Gamma_{14}= 
\frac{1}{2}\big(\Gamma_1^{(-)}+\Gamma_1^{(+)}+\Gamma_2^{(+)}+
\Gamma_3^{(+)}\big) \, , \nonumber \\ 
\Gamma_{32}=\frac{1}{2}\big(\Gamma_2^{(-)}+ \Gamma_3^{(-)}+ 
\Gamma_4^{(-)}+\Gamma_4^{(+)}\big)\, ,\;\;\; \Gamma_{42}= 
\frac{1}{2}\big(\Gamma_1^{(-)}+ \Gamma_3^{(-)}+\Gamma_3^{(+)}+ 
\Gamma_4^{(-)}\big)\, .
\label{e14} \end{eqnarray} 
where labeling of the transitions is indicated in the inset in 
Fig.~1.  Transition rates are: 
\begin{eqnarray}
\Gamma_{1}^{(\pm)}=\mbox{Re} \sum_{j,j'}F_{jj'}(\pm \omega_+)
\sigma_{14}^{(j)}\sigma_{41}^{(j')} \, , \;\;\; 
\Gamma_{2}^{(\pm)}=\mbox{Re} \sum_{j,j'}F_{jj'}(\pm \omega_-)
\sigma_{13}^{(j)}\sigma_{31}^{(j')} \, , \nonumber \\
\Gamma_{3}^{(\pm)}=\mbox{Re} \sum_{j,j'}F_{jj'}(\pm\omega_-) 
\sigma_{24}^{(j)}\sigma_{42}^{(j')}\, , \;\;\;
\Gamma_{4}^{(\pm)}=\mbox{Re}\sum_{j,j'}F_{jj'}(\pm \omega_+) 
\sigma_{23}^{(j)}\sigma_{32}^{(j')} \, . 
\label{e15} \end{eqnarray} 
The superscripts $\pm$ refer here to transitions in the direction 
of decreasing (+) or increasing (-) energy. Finding matrix 
elements $\sigma_{nm}$ from the wavefunctions (\ref{e3}) we see 
explicitly that transitions between the states 1 and 2, as well as 
3 and 4 are suppressed, since the corresponding matrix elements are 
zero, and that the rates (\ref{e15}) are:  
\begin{eqnarray}
\Gamma_{1}= \frac{1}{2} F_{11} (1-\frac{\delta\Delta +\nu^2}{ 
\epsilon \Omega}) +\frac{1}{2} F_{22} (1+\frac{\delta\Delta -\nu^2}
{\epsilon \Omega}) +\mbox{Re} F_{12} (\frac{\nu}{\Omega}-\frac{\nu}
{\epsilon} )\, , \nonumber \\
\Gamma_{2}= \frac{1}{2} F_{11} (1+\frac{\delta\Delta +\nu^2}{ 
\epsilon \Omega}) +\frac{1}{2} F_{22} (1-\frac{\delta\Delta -\nu^2}
{\epsilon \Omega}) +\mbox{Re} F_{12} (\frac{\nu}{\Omega}+\frac{\nu}
{\epsilon}) \, , \nonumber \\
\Gamma_{3}= \frac{1}{2} F_{11} (1+\frac{\delta\Delta +\nu^2}{ 
\epsilon \Omega}) +\frac{1}{2} F_{22} (1-\frac{\delta\Delta -\nu^2}
{\epsilon \Omega}) -\mbox{Re} F_{12} (\frac{\nu}{\Omega}+\frac{\nu}
{\epsilon}) \, , \label{e16}  \\
 \Gamma_{4}= \frac{1}{2} F_{11} (1-\frac{\delta\Delta +\nu^2}{ 
\epsilon \Omega}) +\frac{1}{2} F_{22} (1+\frac{\delta\Delta -\nu^2}
{\epsilon \Omega}) -\mbox{Re} F_{12} (\frac{\nu}{\Omega}-\frac{\nu}
{\epsilon}) \, . \nonumber
\end{eqnarray} 

The transfer ``rates'' $u$, $v$ in Eqs.~(\ref{e12}) and (\ref{e13}) 
are: 
\begin{equation}
u_{\pm}=\sum_{j,j'=1,2}\sigma_{14}^{(j')}\sigma_{32}^{(j)} 
F_{jj'}(\pm\omega_+)\, , \;\;\; 
v_{\pm}=\sum_{j,j'=1,2}\sigma_{13}^{(j')}\sigma_{42}^{(j)}
F_{jj'}(\pm\omega_-) \, .
\label{e17} \end{equation} 
Explicitly: 
\begin{eqnarray}
u= \frac{1}{2} F_{11} (1-\frac{\delta\Delta +\nu^2}{ 
\epsilon \Omega}) -\frac{1}{2} F_{22} (1+\frac{\delta\Delta -\nu^2}
{\epsilon \Omega}) +i\mbox{Im} F_{12} (\frac{\nu}{\Omega}-\frac{\nu}
{\epsilon}) \, ,\nonumber \\
v= -\frac{1}{2} F_{11} (1+\frac{\delta\Delta +\nu^2}{ 
\epsilon \Omega}) +\frac{1}{2} F_{22} (1-\frac{\delta\Delta -\nu^2}
{\epsilon \Omega}) -i\mbox{Im} F_{12} (\frac{\nu}{\Omega}+\frac{\nu}
{\epsilon}) \, .  
\label{e18} \end{eqnarray} 
Equations (\ref{e16}) and (\ref{e18}) do not show the frequency 
dependence of noise correlators, which is the same, respectively, as 
in the Eqs.~(\ref{e15}) and (\ref{e17}). 

Each pair, (\ref{e12}) and (\ref{e13}), of coupled equations can be 
solved directly by diagonalization of the matrix of the evolution 
coefficients with a non-orthogonal transformation. In this way we 
obtain for the pair of equations (\ref{e12}): 
\begin{eqnarray}
\rho_{13}(t)= \frac{1}{u_+u_-+c^2}\big[ \rho_{13}(0)(u_+u_-e^{- 
\gamma_+t}+c^2e^{- \gamma_-t}) +cu_-\rho_{42}(0)(e^{-\gamma_+t}- 
e^{-\gamma_-t}) \big]\, ,\nonumber \\
\rho_{42}(t)= \frac{1}{u_+u_-+c^2}\big[ \rho_{42}(0)(u_+u_-e^{- 
\gamma_-t}+c^2e^{- \gamma_+t}) +cu_+\rho_{13}(0)(e^{-\gamma_+t}- 
e^{-\gamma_-t})\big]\, . 
\label{e19} \end{eqnarray} 
where 
\begin{equation} 
\gamma_{\pm} \equiv (\Gamma_{13}+\Gamma_{42})/2 \pm \big[
(\Gamma_{13}-\Gamma_{42})^2/4+u_+u_-\big]^{1/2} \, , \;\;\;
c\equiv (\Gamma_{13}-\Gamma_{42})/2 - \big[ (\Gamma_{13}- 
\Gamma_{42})^2/4+u_+u_-\big]^{1/2} \, , 
\label{e20} \end{equation}
and $\rho_{13}(0)$, $\rho_{42}(0)$ are the initial values 
of the density matrix elements that depend on preparation of 
the initial state. If, as in the experiment \cite{b7}, the qubits 
are abruptly driven to co-resonance maintaining the state 
$|00\rangle$, these initial values are: 
\begin{eqnarray}
\rho_{13}(0)=\frac{1}{4}(\gamma+\eta)(\alpha+\beta)\, , \;\;\; 
\rho_{32}(0)=\frac{1}{4}(\gamma-\eta)(\alpha+\beta)\, , \nonumber
\\ \rho_{14}(0)=\frac{1}{4}(\gamma+\eta)(\beta-\alpha)\, , \;\;\; 
\rho_{42}(0)=\frac{1}{4}(\gamma-\eta)(\alpha-\beta)\, . 
\label{e21} \end{eqnarray} 
Another type of initial conditions that will be discussed in this 
work is starting the oscillations from the state $|10\rangle$. In 
this case: 
\begin{eqnarray}
\rho_{13}(0)=\frac{1}{4}(\gamma-\eta)(\alpha-\beta)\, , \;\;\; 
\rho_{32}(0)=\frac{1}{4}(\gamma+\eta)(\alpha-\beta)\, ,\nonumber
\\ \rho_{14}(0)=\frac{1}{4}(\gamma-\eta)(\alpha+\beta)\, , \;\;\;  
\rho_{42}(0)=\frac{1}{4}(\gamma+\eta)(\alpha+\beta)\, . 
\label{e22} \end{eqnarray} 
Equations (\ref{e21}) and (\ref{e22}) follow directly from the 
wavefunctions (\ref{e3}): $\rho_{nm}(0)=\langle n|i\rangle \langle 
i|m\rangle$, where $|i\rangle$ is the initial state. 

Solution of the other pair (\ref{e13}) of coupled equation is given 
by the same Eqs.~(\ref{e19}) and (\ref{e20}) with obvious 
substitutions: $u_{\pm} \rightarrow v_{\pm}$, $\Gamma_{13} 
\rightarrow \Gamma_{14}$, $\Gamma_{42} \rightarrow \Gamma_{32}$. 
In this work, we are mostly 
interested in the low-temperature regime $T\ll \epsilon, \Omega$, 
when transitions  up in energy can be neglected. In this regime, 
$u_-,v_- \rightarrow 0$, and equations for the 
evolution of the density matrix elements are simplified. For 
instance, for $u_- \rightarrow 0$, $c\simeq u_+u_-/(\Gamma_{13}- 
\Gamma_{42})$, and Eqs.~(\ref{e19}) are reduced to:   
\begin{equation}  
\rho_{13}(t)= \rho_{13}(0)e^{-\Gamma_{13}t}\, ,\;\;\; 
\rho_{42}(t)= \rho_{42}(0)e^{-\Gamma_{42}t}+\frac{u_+}{\Gamma_{13}- 
\Gamma_{42}}\rho_{13}(0)(e^{-\Gamma_{13}t}-e^{-\Gamma_{42}t}) \, ,  
\label{e23} \end{equation} 
where now $\Gamma_{13}=(\Gamma_1+ \Gamma_2+ \Gamma_4)/2$ and 
$\Gamma_{42}= \Gamma_3/2$. 

Time evolution (\ref{e23}) of the density matrix elements together 
with the rates (\ref{e16}) and (\ref{e18}), and initial conditions 
(\ref{e21}) and (\ref{e22}) determines the shape of the coherent 
oscillations in two coupled qubits. In the next Section, we discuss 
this shape in several specific situations.

\section{Results and Conclusions}

The shape of coherent oscillations determined in the previous 
Section depends on the large number of parameters: temperature, 
degree of asymmetry of qubit tunnel energies and couplings to the 
environments, frequency dependence of the decoherence, and strength 
and nature of the correlations between the two reservoirs. We 
analyze some of these dependencies below.  

\subsection{Experimentally-motivated case} 

We start by considering the situation that is close to the 
experimentally realized case of oscillations in coupled charge 
qubits \cite{b7}. As we argued above, the correlations between 
environments 
in this case should be real: $\mbox{Im} F_{12}=0$. The oscillations 
are excited by driving the system to co-resonance in the initial 
state $|00\rangle$. Equations of the previous Section give in this 
case the following expression for the shape of the oscillations: 
\begin{eqnarray} 
p_{1}(t)=\frac{1}{2}-\frac{1}{8}\big[A e^{-\Gamma_{42}t} +B 
(e^{-\Gamma_{13}t}-\frac{u_+}{\Gamma_{13}- \Gamma_{42}}
(e^{-\Gamma_{13}t}-e^{-\Gamma_{42}t}))\big] \cos \omega_-t 
\nonumber \\
-\frac{1}{8}\big[C e^{-\Gamma_{32}t} +D 
(e^{-\Gamma_{14}t}+\frac{v_+}{\Gamma_{14}- \Gamma_{32}}
(e^{-\Gamma_{14}t}-e^{-\Gamma_{32}t}))\big] \cos \omega_+t \, ,
\label{e24} \end{eqnarray} 
where 
\[ A=1+\frac{\delta\Delta +\nu^2}{ 
\epsilon \Omega}- \frac{\nu}{\Omega}-\frac{\nu}{\epsilon} \, , \;\;\; 
B=1+\frac{\delta\Delta +\nu^2}{\epsilon \Omega} +\frac{\nu}{\Omega}+ 
\frac{\nu}{\epsilon} \, , \;\;\; 
C=1-\frac{\delta\Delta +\nu^2}{ \epsilon \Omega}- \frac{\nu}{\Omega}+ 
\frac{\nu}{\epsilon} \, , \;\;\;
D=1-\frac{\delta\Delta + \nu^2}{\epsilon \Omega} +
\frac{\nu}{\Omega}- \frac{\nu}{\epsilon} \, . \]

Equation for $p_2(t)$ is the same with signs in front of $\delta$, 
$u_+$ and $v_+$ reversed. As a simplifying assumption we take 
$F_{11}=F_{22}\equiv F$. The decoherence rates in Eqn.~(\ref{e24}) 
then are:
\begin{eqnarray}
\Gamma_{13}= F(\omega_+) (1-\frac{\nu^2}{ \epsilon \Omega}) 
+\frac{1}{2} F(\omega_-) (1+\frac{\nu^2}{\epsilon \Omega}) 
+\frac{1}{2}F_c(\omega_-) (\frac{\nu}{\Omega}+\frac{\nu}
{\epsilon} ) \, , \;\;\; u_+= - F(\omega_+) \frac{\delta 
\Delta}{\epsilon \Omega} \, , \nonumber \\
\Gamma_{14}= F(\omega_-) (1+\frac{\nu^2}{ \epsilon \Omega}) 
+\frac{1}{2} F(\omega_+) (1-\frac{\nu^2}{\epsilon \Omega}) 
+\frac{1}{2}F_c(\omega_+) (\frac{\nu}{\Omega}-\frac{\nu}
{\epsilon}) \, , \;\;\; v_+= - F(\omega_-) \frac{\delta 
\Delta}{\epsilon \Omega}\, , \label{e25} \\
\Gamma_{32}= \frac{1}{2} F(\omega_+) (1-\frac{\nu^2}{\epsilon 
\Omega}) +\frac{1}{2} F_c (\omega_+) (\frac{\nu}{\epsilon}- 
\frac{\nu}{\Omega}) \, , \;\;\; 
\Gamma_{42}= \frac{1}{2} F(\omega_-) (1+\frac{\nu^2}{\epsilon 
\Omega}) -\frac{1}{2} F_c (\omega_-) (\frac{\nu}{\epsilon}+ 
\frac{\nu}{\Omega}) \, , \nonumber 
\end{eqnarray} 
where $F_c(\omega)\equiv \mbox{Re} F_{12}(\omega)$. To enable 
comparison of these rates to those of individual qubits, we note that 
the rate of decoherence of oscillations in a single qubit  
with vanishing bias is equal to $F(\Delta_j)/2$ for the $j$th qubit.

The functions $p_{1,2}(t)$ for the qubit parameters, $\delta$, $\nu$, 
and $F$, close to those in experiment \cite{b7} are plotted in Fig.\ 
2 under additional assumption that the decoherence is the same at 
two frequencies, $\omega_+$ and $\omega_-$. (The decoherence strength 
$F$ was taken from the data for the single-qubit regime in \cite{b7}.) 
The curves are plotted for the two situations, when decoherence is 
completely uncorrelated ($F_{12}=0$) and completely correlated 
($F_{12}=F$) between the two qubits. One can see that the difference 
between the two regimes is not very big numerically, with 
correlations between the two reservoirs leading to the effective 
decoherence rate that is increased in comparison with the uncorrelated 
regime by roughly $30\div50$\%, although the description with a single 
decoherence rate is not quite appropriate quantitatively -- see 
Eqs.~(\ref{e24}) and (\ref{e25}). 

\begin{figure}[htb]
\setlength{\unitlength}{1.0in}
\begin{picture}(5.0,3.3) 
\put(1.4,-.1){\epsfxsize=3.4in\epsfbox{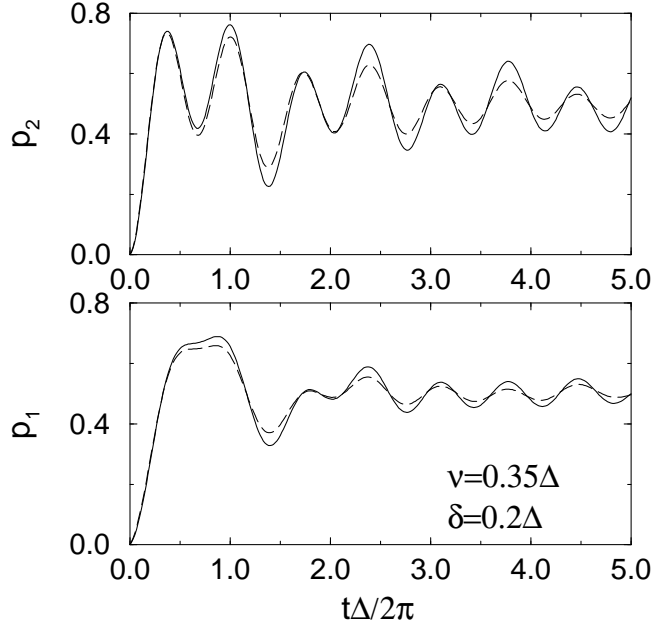}}
\end{picture}
\caption{Probabilities $p_j$ to find $j$th qubit in the state 
$|1\rangle$ in the process of quantum coherent oscillations 
starting with the state $|00\rangle$ of two coupled qubits. 
The decoherence strength is $F=0.08\Delta$. Solid and dashed 
lines correspond, respectively, to the decoherence that is 
uncorrelated ($F_c=0$) and completely correlated ($F_c=F$) 
between the two qubits. }
\end{figure}

The increase of the effective decoherence rate by correlations 
illustrated in Fig.\ 2 can be related to the fact that the 
initial qubit state, $|00\rangle$, belongs to the subspace where 
the correlations increase the decoherence rate, despite the mixing 
of this subspace with the DFS where the decoherence rate is 
decreased in the eigenstates (\ref{e3}) of the coupled qubit 
system. (Here and below we use the term ``DFS'' for the 
subspace spanned by the $|01\rangle$ and $|10\rangle$ states, 
although for interacting qubits it, strictly speaking, does not 
fully have the properties of real DFS.) This implies that increase 
of decoherence rate by correlations should be to a large extent 
insensitive to qubit parameters. This conclusion is supported 
by the case of identical qubits ($\delta =0$), when $u_+=v_+=0$ 
and Eq.~(\ref{e24}) is reduced to a very simple form: 
\begin{equation} 
p_{1}(t)=\frac{1}{2}-\frac{1}{4}\big[(1+\frac{\nu}{ \Omega})
e^{-\Gamma_{13}t} \cos \omega_-t  + (1- \frac{\nu}{\Omega}) 
e^{-\Gamma_{32}t} \cos \omega_+t  \big]\, , 
\label{e26} \end{equation} 
and $p_2(t)=p_{1}(t)$. One can see from Eqs.~(\ref{e25}) 
that both decoherence rates relevant for Eq.~(\ref{e26}), 
$\Gamma_{13}$ and $\Gamma_{32}$ increase with increasing 
correlation strength $F_c$. Equation (\ref{e26}) shows also 
that the description of the oscillation decay with a single 
decoherence rate can be quite inaccurate: for weak interaction, 
$\nu < \Omega$ the amplitudes of the two (high- and low-frequency) 
components of the oscillations are nearly the same while their 
decoherence rates can be very different.

\subsection{Excitation into the DFS} 

Now we discuss decoherence properties of the oscillations in 
coupled qubits in the case when they start with the initial qubit 
state $|10\rangle$. We note that in the case of experiment similar 
to \cite{b7}, such an initial condition would require separate 
gate control of the two qubits, since the bias change bringing 
them into co-resonance is different in this state for the two 
qubits. Since the state $|10\rangle$ belongs to the DFS in the 
case of completely correlated noise, one can expect that 
oscillations with these initial conditions will be more sensitive 
to the degree of inter-qubit decoherence correlations than 
oscillations with $|00\rangle$ initial condition, and that the 
effective decoherence rate will decrease with correlation strength. 
All this indeed can be seen from Eqs.~(\ref{e23}) with the 
initial conditions (\ref{e22}) that correspond to the $|10\rangle$ 
state. Under the same assumptions as were used in Eqn.~(\ref{e24}),  
we get for the  now different $p_1(t)$ and $p_2(t)$: 
\begin{eqnarray} 
p_{j}(t)=\frac{1}{2}-\frac{(-1)^j}{8}\big\{ \big[A_j e^{-\Gamma_{42} 
t} +B_j (e^{-\Gamma_{13}t}+\frac{(-1)^ju_+}{\Gamma_{13}- \Gamma_{42}}
(e^{-\Gamma_{13}t}-e^{-\Gamma_{42}t}))\big] \cos \omega_-t 
\nonumber \\
+ \big[C_j e^{-\Gamma_{32}t} +D_j (e^{-\Gamma_{14}t}-\frac{(-1)^j 
v_+}{\Gamma_{14}- \Gamma_{32}} (e^{-\Gamma_{14}t}-e^{-\Gamma_{32}t})) 
\big] \cos \omega_+t \big\} \, , \;\;\; j=1,2 \, , 
\label{e27} \end{eqnarray} 
where 
\[ A_1=1+\frac{\delta\Delta +\nu^2}{ 
\epsilon \Omega}+ \frac{\nu}{\Omega}+\frac{\nu}{\epsilon} \, , 
\;\;\; B_1=1+\frac{\delta\Delta +\nu^2}{\epsilon \Omega} - 
\frac{\nu}{\Omega} - \frac{\nu}{\epsilon} \, , \] 
\[ C_1=1-\frac{\delta\Delta +\nu^2}{ \epsilon \Omega} + 
\frac{\nu}{\Omega} -\frac{\nu}{\epsilon} \, , \;\;\;
D_1=1-\frac{\delta\Delta + \nu^2}{\epsilon \Omega} -
\frac{\nu}{\Omega}+ \frac{\nu}{\epsilon} \, , \]
and the amplitudes $A_2,\, B_2\, , C_2\, , D_2$ are given by the 
same expressions with $\delta \rightarrow - \delta$. 

For identical qubits Eqn.~(\ref{e27}) reduces to: 
\begin{equation} 
p_{j}(t)=\frac{1}{2}-\frac{(-1)^j}{4}\big[(1+\frac{\nu}{ \Omega})
e^{-\Gamma_{42}t} \cos \omega_-t + (1- \frac{\nu}{\Omega}) 
e^{-\Gamma_{14}t} \cos \omega_+t  \big]\, .  
\label{e28} \end{equation} 
This expression and Eqs.~(\ref{e25}) show that in contrast 
to Eqn.~(\ref{e26}), the decoherence rate of the low-frequency 
component that has larger amplitude is strongly suppressed by 
the non-vanishing inter-qubit noise correlations $F_c$: 
$\Gamma_{42}= \frac{1}{2} (1+\nu/ \Omega) [F(\omega_-) - 
F_c (\omega_-) ]$. This means that the shape of the coherent 
oscillations in coupled qubit starting with the state $|10\rangle$ 
should indeed be more sensitive to the strength of these 
correlations than the shape of the oscillations starting with 
the $|00\rangle$ state. As one can see from Fig.\ 3 which 
shows the shape (\ref{e27}) of the $|10\rangle$ oscillations for 
the same set of experimentally realized parameters as in Fig.\ 3, 
this conclusion also remains valid in the case of not fully 
symmetric qubits. Even in this case there is a pronounced weakly 
decaying component of the oscillations if the decoherence is 
completely correlated between the two qubits. For partial 
correlations, the effective decoherence rate is reduced. 

\begin{figure}[htb]
\setlength{\unitlength}{1.0in}
\begin{picture}(5.0,3.5) 
\put(1.3,-.1){\epsfxsize=3.6in\epsfbox{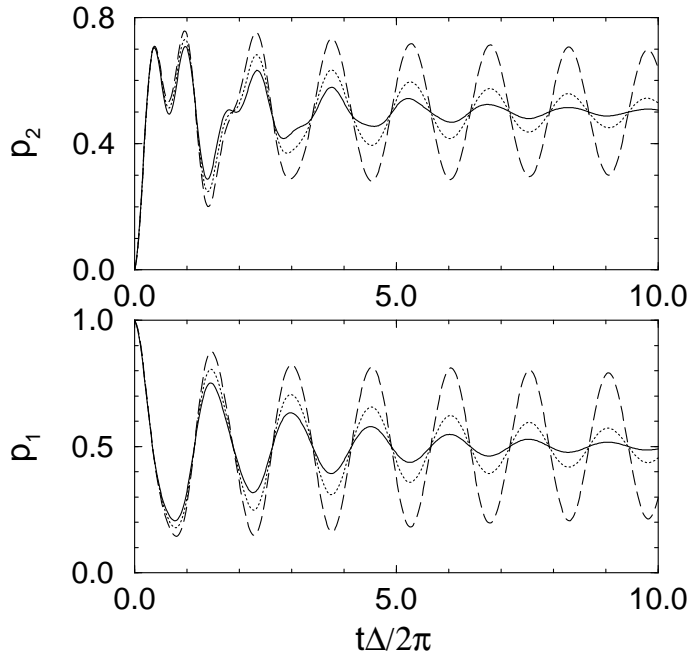}}
\end{picture}
\caption{Probabilities $p_j$ to find $j$th qubit in the state 
$|1\rangle$ in the process of quantum coherent oscillations 
starting with the state $|10\rangle$ of two coupled qubits. 
Qubit parameters are the same as in Fig.\ 2. Solid, dotted, and 
dashed lines correspond, respectively, to the decoherence that 
is uncorrelated ($F_c=0$), partially ($F_c=0.5F$), and completely 
($F_c=F$) correlated between the two qubits. }
\end{figure}

In summary, we have developed quantitative description of 
weakly dissipative dynamics of two coupled qubits based on the 
standard Markovian evolution equation for the density matrix. 
This description shows that decoherence properties of currently 
realized oscillations in coupled qubits are not very sensitive 
to inter-qubit correlations of decoherence, while relatively 
simple modification of the excitation scheme for the oscillations 
should make them sensitive to these correlations.

\section*{Acknowledgements} 

D.V.A. would like to thank RIKEN and NEC Basic Research 
Laboratories for their hospitality during a visit when part of 
this work was done, and the group of Prof. F. Nori for 
discussions that stimulated it. The authors acknowledge useful 
discussions with O. Astafiev, J.E. Lukens, Y. Nakamura, 
Yu.A. Pashkin, J.S. Tsai, and T. Yamamoto. This work was supported 
in part by ARDA and DOD under the DURINT grant \# F49620-01-1-0439 
and by the NSF under grant \# 0121428.

\end{document}